\documentclass[11pt]{article}

\usepackage{xcolor}
\usepackage{ulem}
\usepackage{eps fig}
\usepackage{amsfonts}
\usepackage{latexsym}
\usepackage{amsmath,amssymb,bm}
\usepackage{verbatim}
\usepackage{color}
\usepackage{hyperref}
\usepackage{cite}
\usepackage{slashed}

\usepackage{setspace}
\usepackage[textheight=9in, textwidth=6.5in, letterpaper]{geometry}

\def\half{{1\over 2}}
\numberwithin{equation}{section}

 \def\p{\partial}

\newcommand{\bea}{\begin{eqnarray}}
\newcommand{\eea}{\end{eqnarray}}
\newcommand{\be}{\begin{equation}}
\newcommand{\ee}{\end{equation}}
\newcommand{\ba}{\begin{align}}
\newcommand{\ea}{\end{align}}

\newcommand{\Pol}{\text{Pol}(\Delta)}

  \makeatletter
  \let\over=\@@over \let\overwithdelims=\@@overwithdelims
  \let\atop=\@@atop \let\atopwithdelims=\@@atopwithdelims
  \let\above=\@@above \let\abovewithdelims=\@@abovewithdelims

\renewcommand\section{\@startsection {section}{1}{\z@}%
                                   {-3.5ex \@plus -1ex \@minus -.2ex}
                                   {2.3ex \@plus.2ex}%
                                   {\normalfont\large\bfseries}}

\renewcommand\subsection{\@startsection{subsection}{2}{\z@}%
                                     {-3.25ex\@plus -1ex \@minus -.2ex}%
                                     {1.5ex \@plus .2ex}%
                                     {\normalfont\bfseries}}

\newcommand{\Tr}{\mbox{Tr}}

\def\half{{1 \over 2}}

\def\Or[#1]{{\text{O}}\left({#1}\right)}
\def\dotl[#1,#2]{\left\langle #1, #2 \right\rangle}
\def\dotlb[#1,#2]{[ #1, #2 ]}
\def\dotp[#1,#2]{(#1) \cdot (#2)}
\def\aff[#1,#2]{\hat{#1}(#2)}
\def\n4sym{{\cal N}=4 SYM}
\def\>{\rangle}
\def\<{\langle}

\def\weight[#1,#2,#3]{\{(#1),#2,#3\}}
\def\ads[#1]{$\text{AdS}_{#1}$}

\newcommand{\nn}{\nonumber}

\newcommand{\beq}{\begin{equation}}
\newcommand{\eeq}{\end{equation}}
\newcommand{\D}{\Delta}

\def\hstate{\left| h \right> }
\def\hstate{\left| h \right> }
\def\hbra{\left< h \right| }
\def\Lads{\ell_A}

\def\calL{{\cal L}}
\def\fA{{\bf A}}

\begin{document}

\unitlength = 1mm

\pagestyle{empty}

\setcounter{page}{0}
\pagestyle{plain}

\setcounter{tocdepth}{3}

\def\vx{{\vec x}}
\def\ip{${\cal I}^+$}
\def\p{\partial}
\def\po{$\cal P_O$}
\def\hstatespin{\left| h,a \right> }
\def\hstatespinb{\left| h,b \right> }

\thispagestyle{empty}
\begin{center}
{  \Large{\textsc{Partition Functions with spin in AdS$_2$ \\via Quasinormal Mode Methods}} }
\vspace*{0.25cm}

Cynthia Keeler$^{a}$,
Pedro Lisb\~ao$^{b}$
and Gim Seng Ng$^{c}$
\vspace*{0.25cm}

{\it $^a$Niels Bohr International Academy, Niels Bohr Institute
University of Copenhagen, Blegdamsvej 17, DK 2100, Copenhagen, Denmark}

{\it $^b$Department of Physics, University of Michigan,
Ann Arbor, MI-48109, USA}

{\it $^c$Department of Physics, McGill University, Montr\'eal, QC H3A 2T8, Canada.}

\end{center}

\begin{abstract}
We extend the results of \cite{Keeler:2014hba}, computing one loop partition functions for massive fields with spin half in AdS$_2$ using the quasinormal mode method proposed by Denef, Hartnoll, and Sachdev \cite{Denef:2009kn}. We find the finite representations of $SO(2,1)$ for spin zero and spin half, consisting of a highest weight state $|h\rangle$ and descendants with non-unitary values of $h$. 
These finite representations capture the poles and zeroes of the one loop determinants. Together with the asymptotic behavior of the partition functions (which can be easily computed using a large mass heat kernel expansion), these are sufficient to determine the full answer for the one loop determinants.
We also discuss extensions to higher dimensional AdS$_{2n}$ and higher spins.
\end{abstract}
\newpage
\tableofcontents

\section{Introduction}

The computation of quantum fluctuations in AdS spacetimes is of great interest due to their role in the AdS$_{d+1}$/CFT$_{d}$ correspondence \cite{Gubser:2002zh,Gubser:2002vv,Hartman:2006dy,Diaz:2007an,Diaz:2008iv,Giombi:2008vd,Aros:2009pg,Denef:2009yy,Denef:2009kn,Faulkner:2013ana,Barrella:2013wja,Giombi:2013fka,Giombi:2014iua,Giombi:2014yra} and their relationship to the microscopic structure of extremal black holes \cite{Sen:2008vm,Banerjee:2010qc,Mandal:2010cj,Bhattacharyya:2012ye}. The leading quantum correction to the entropy of an extremal black hole is evaluated via the functional determinant or one loop effective action. 

One well-studied method for computing these one loop determinants is the heat kernel \cite{Vassilevich:2003xt}. The heat kernel can be expanded at large mass for any spacetime (and many operators) in terms of a set of local curvature scalars.  However, in order to compute the full one loop determinant, we need to know the spectra of all fluctuating fields present, as well as their eigenfunctions. This method is well-established but becomes unwieldy for high numbers of fields or non-minimal couplings, because those cases involve complicated diagonalization of the mass matrices. Simplifications can be made if the group theory structure of the field content is well-understood \cite{David:2009xg,Gopakumar:2011qs}. Recently, \cite{Larsen:2014bqa,Keeler:2014bra,Larsen:2015aia} worked towards streamlining the calculation of determinants relying on the on-shell spectrum but still using the heat kernel method.

In \cite{Denef:2009kn}, Denef, Hartnoll, and Sachdev developed a fundamentally different approach for calculating the one loop determinant, by studying it in the complex mass plane. Instead of computing the partition function $Z(\Delta)$ for a given mass set by the conformal dimension $\Delta$, they treat $Z(\Delta)$ as a function on the complex $\Delta$ plane. If $Z(\Delta)$ is a meromorphic function in the complex $\Delta$ plane - which we will see is a reasonable assumption for our purposes - one can determine the function from the location and multiplicities of its poles and zeros, up to a polynomial function of $\Delta$. We fix the polynomial part by studying the large mass behavior of $Z(\Delta)$ via the local curvature expansion of the heat kernel method.

In \cite{Keeler:2014hba}, we used this quasinormal mode method to compute partition functions for scalars in even dimensional AdS$_{2n}$ spaces. In this note, we apply this method to massive fields with spin half, spin one, and spin two in the context of AdS$_2$.\footnote{For applications of the quasinormal mode method to odd-dimensional AdS, we refer the readers to \cite{Datta:2011za,Zhang:2012kya} in the case of AdS$_3$.} 
We connect the modes responsible for the poles or zeroes in the one loop determinant to finite representations of $SO(2,1)$. This connection further simplifies calculation of the one loop determinant.

In Sec.~\ref{sec:reviewDHS} we review the quasinormal mode method of Denef, Hartnoll, and Sachdev (hereafter DHS) \cite{Denef:2009kn} and the results of \cite{Keeler:2014hba} for massive scalars in AdS$_2$.
In Sec.~\ref{sec:spin0}, we compute the modes responsible for partition function poles of the massive scalar field via finite representations of $SO(2,1)$, efficiently reproducing the modes previously found in \cite{Keeler:2014hba}.
In Sec.~\ref{sec:spinhalf}, we extend this algebraic method to effortlessly compute the relevant spin half (Dirac spinor) modes and reproduce known results in the literature.  In Sec.~\ref{sec:spinoneandtwo}, we discuss how to use the algebraic method to generalize the computations to spin one and two in AdS$_2$. Along the way, we also discuss generalizations of the spin zero and spin half (Dirac spinor) in higher even-dimensional AdS$_{2n}$, with details given in Appendix~\ref{sec:app1}.


\section{Computing zero modes}
\label{sec:reviewDHS}
In this section we review the  DHS method \cite{Denef:2009kn} for computing one loop determinants and partition functions. We will focus on a {\it complex} scalar field in AdS$_2$ (with AdS length $\Lads$) as a guiding example, following the computation in \cite{Keeler:2014hba}.

The central idea of the DHS method is to consider the one loop determinant as a function of a mass parameter, and then continue that mass parameter to the complex plane.  Considering the determinant as a function of a complex mass parameter allows us to use the power of complex analysis, in particular Weierstrass's factorization theorem. This theorem states that any meromorphic function on the complex plane can be determined from its zeroes and poles.\footnote{More precisely, we mean the meromorphic extension of Weierstrass's factorization theorem.  Additionally, as detailed below, the poles and zeros only determine a meromorphic function up to one overall function, which itself cannot have any zeros or poles (that is, up to one {\it entire} function).} We are specifically interested in computing one loop determinants, whose zeroes and poles in the complex mass plane can be found from the kinetic operator's spectrum.
Hence Weierstrass's theorem provides a shortcut for calculating one loop determinants, provided we assume they are meromorphic.

Let us consider the example of the complex scalar field in AdS$_2$. At one loop, its partition function is proportional to the inverse determinant of the massive Klein-Gordon operator,
\be\label{eq:partifunctionscalar}
Z(\D) = \int \mathcal{D}\phi e^{-\int \phi^* [-\nabla^2 + \Lads^{-2}\D(\D-1)]\phi} \propto \frac{1}{\det[-\nabla^2 + \Lads^{-2}\D(\D-1)]}.
\ee
The conformal dimension $\D$ can be expressed in terms of the mass $m$ of the complex scalar via $\D(\D-1) = (m\ell_A)^2$, or equivalently,
\be \label{eq:deltaandm}
\D = \frac{1}{2} + \sqrt{\frac{1}{4} + m^2\ell_A^2}.
\ee
Since the boundary conditions in AdS spaces are usually defined in terms of the conformal dimension $\Delta$, we will continue this parameter (rather than $m$) to the complex plane. 

By inspection of Eq.~(\ref{eq:partifunctionscalar}) we see that the partition function $Z$ is a function of $\D$ with no zeroes and with poles located at $\D=\D_{\star}$, where $\D_{\star}$ is a particular value of the conformal dimension for which there exists a $\phi_{\star}$ satisfying 
\be \label{eq:scalarpoles}
[-\nabla^2 + \ell_A^{-2}\D_{\star}(\D_{\star}-1)]\phi_{\star} =0.
\ee
That is, $\phi_{\star}$ is a zero mode of the Klein-Gordon operator with mass set by the conformal dimension $\D_\star$.  In order for $\D_\star$ to indicate a pole in $Z(\D)$, its associated solution $\phi_{\star}$  must be single-valued and contain only the ``normalizable'' behavior at the conformal boundary of AdS$_2$. In global coordinates
\be\label{eq:etathetametric}
ds^2 = \ell_A^2 (d\eta^2 + \sinh^2\eta d\theta^2),~ \theta\sim\theta+2\pi,~\eta\geq0,
\ee
the boundary and single-valued conditions on the solutions $\phi_{\star}$ become
\begin{align}\label{eq:bcandperiodic}
&\phi_\star \rightarrow (\sinh \eta)^{-\D}~~\textrm{when}~~\eta\rightarrow \infty, \\
&~~~~~~~~ \phi_\star(\theta) = \phi_\star(\theta+2\pi). 
\end{align}
For AdS$_2$, the explicit solutions $\phi_\star$ are given in \cite{Keeler:2014hba}:
\begin{align}
\nn
\phi_{hl} &= 
e^{il\theta}(i\sinh \eta)^{| l |} F\left[h+|l|, |l| + 1 - h; |l| +1; -\sinh^2\left(\frac{\eta}{2}\right) \right],
\\\label{eq:explicitphistar}
h & \in {\mathbb Z}_{\leq 0}, ~  l\in {\mathbb Z}, ~ |l| \leq -h.
\end{align}
These $\phi_\star$ solve Eq. (\ref{eq:scalarpoles}) under the boundary conditions (\ref{eq:bcandperiodic}), with $\D_\star = h$.

There are $-2h+1$ solutions $\phi_\star$ for each value of $h=\D_\star$, we denote this degeneracy by $D_h$. The partition function $Z(\D)$ is then given by
\be \label{eq:weierstrass}
Z(\D) = e^{\text{Pol}(\D)}\prod_h\frac{1}{(\D - h)^{D_{h}}}.
\ee
Here $h$ is an index labelling the distinct poles $\D_\star=h$ of $Z(\D)$, with the product running over all nonpositive integers $h \in {\mathbb Z}_{\leq 0}$. $\text{Pol}(\D)$ is an as-yet undetermined function; it must be polynomial in $\D$ since it cannot contribute any new poles or zeros to $Z(\D)$.

Rather than continue to work with an infinite product, we take the logarithm of $Z(\D)$:
\begin{align} \label{eq:logZ}
\log Z(\D) &=
 \text{Pol}(\D) - \sum_h D_h \log(\D - h),
\\\label{eq:logZDelta}
&= \text{Pol}(\D) + 2 \zeta'(-1,\D)- (2\D-1)\zeta'(0,\Delta).
\end{align}
where in the second line we have treated the infinite sum via zeta function regularization.%
\footnote{$\Pol$ is sufficient to account for any zeta function regularization ambiguities in the cases we study; we expect this behavior to be generic.} %
$\zeta(s,x)$ is the Hurwitz zeta function, found by the analytic continuation of $\zeta(s,x)=\sum_{k=0}^\infty(x+k)^{-s}$, and $\zeta'(s,x) = \partial_s \zeta(s,x)$.

The only undetermined part of the partition function $Z(\D)$ at this point is the polynomial $\Pol$. This polynomial encodes the behavior of $Z(\D)$ at large $\D$, which can be computed from a large mass heat kernel expansion, where $\D$ and $m$ are related by (\ref{eq:deltaandm}). The heat kernel expansion of $Z(\D)$ at large $m$ (and thus large $\Delta$) for a generic spacetime of dimension $d+1$ is given by \cite{Vassilevich:2003xt},
\be \label{eq:largemassheatkernel}
\log Z(\D) = \sum_{k=0}^{d+1} a_k \int_0^\infty \frac{dt}{t} t^{\frac{k-(d+1)}{2}}e^{-tm^2}+ \mathcal{O}(m^{-1}) + \textrm{constant}.
\ee
The coefficients $a_k$ encode information about the operator in the one loop determinant as well as the manifold geometry and background fields; they are given by combinations of curvature invariants such as $R, R_{\mu\nu},$ etc.%
\footnote{In the presence of a background gauge field the $a_k$ would also have insertions of the field strength $F_{\mu\nu}$.} %
In our current example, with $d+1=2$ and the Klein-Gordon operator, the nonzero coefficients are\footnote{For an even dimensional spacetime with no boundary contribution all the odd $k$ coefficients are zero.} 
\begin{equation}
a_0 = \frac{1}{(4\pi)} \Tr \int_{{\rm AdS}_2}\sqrt{g}d^2x, \qquad a_2= \frac{1}{(4\pi)} \Tr \int_{{\rm AdS}_2}\sqrt{g}d^2x\frac{R}{6}.
\end{equation}
The integrals over the manifold yield factors of the regularized volume of AdS$_{2}$, since $R = -\frac{2}{\ell_A^2}$ is a constant. The trace in the definition of the $a_k$ sums over the Lorentz index structure of the fields, which is trivial for a scalar. With these coefficients, and using the regulated Vol$_{{\rm AdS}_2} = -2\pi\ell_A^2$, the heat kernel expansion for a scalar in AdS$_2$ is
\be
\log Z(\Delta) = -\frac{\ell_A^2}{2} \int_\epsilon^\infty dt \bigg( \frac{1}{t^2} -\frac{1}{3t\ell_A^2} \bigg)e^{-tm^2}+ \mathcal{O}(m^{-1}) + \textrm{constant}.
\ee
Evaluating this integral with cutoff $\epsilon=e^{-\gamma}\Lambda^{-2}$ determines the large mass, or large $\Delta$, behavior of $Z(\D)$.  As shown explicitly in \cite{Keeler:2014hba}, requiring Eq. (\ref{eq:logZDelta}) to match this large $\D$ behavior fixes $\Pol$:
\beq
\Pol = [-1+\log(\ell_A \Lambda)]\D(\D-1)+\frac{1}{3}\log(\ell_A \Lambda) -\frac{1}{4}.
\eeq
And we now have an expression for the partition function at any $\D$:
\beq
\log Z(\D) = 2\zeta'(-1,\Delta)-(2\Delta-1)\zeta'(0,\Delta)+ [-1+\log(\ell_A \Lambda)]\D(\D-1)+\frac{1}{3}\log(\ell_A \Lambda) -\frac{1}{4}.
\eeq
It is important to note that we only needed the large mass expansion (\ref{eq:largemassheatkernel}), instead of the full heat kernel. 

In summary, we have outlined in this section the zero-mode or DHS method \cite{Denef:2009kn}  for computing partition functions. We have also reviewed the specific case of the AdS$_2$ scalar partition function, as computed in \cite{Keeler:2014hba} for the broader case of AdS$_{2n}$, by following the prescription: 
\begin{itemize}
\item Find all zero modes $\phi_\star$ as well as their conformal dimensions $\Delta_\star$ and degeneracies $D_\star$.
\item Use zeta function regularization to write the logarithm of the partition function.
\item Match the asymptotic behavior of $Z(\D)$ with that of the heat kernel curvature expansion to find $\text{Pol}(\D)$.
\end{itemize}

The evaluation of one loop determinants for other fields follows the prescription outlined here for the scalar field.


\section{An algebraic approach to scalar zero modes}
\label{sec:spin0}

In \cite{Keeler:2014hba}, we found the modes $\phi_\star$ that solve (\ref{eq:scalarpoles}) by explicitly solving the equation of motion and finding values of $\D_\star$ for which $\phi_\star$ had the desired boundary behavior.  In this section, we will show that these same modes can be produced from studying the highest weight representations for the $SO(2,1)$ isometry group of AdS$_2$, even though these representations will not be unitary under the $L^2$ norm.  We then explain how to generalize this algebraic method to higher dimensional AdS$_{2n}$.

\subsection{The $SL(2,R)$ algebra}

We will find it useful to consider the $SL(2,R)$ algebra which is isomorphic to the isometry group $SO(2,1)$ of AdS$_2$.  The algebra is generated by $L_0, \, L_{\pm1}$, and satisfies the commutation relations
\begin{align}
\left[L_0, L_\pm\right] =\mp L_\pm,\quad
\left[L_+, L_-\right] & = 2L_0,
\end{align}
where we have abbreviated $L_{\pm 1}=L_\pm$.
The quadratic Casimir for this algebra is $L_0^2  - L_0 -L_{-}L_{+}$.  States with well-defined conformal dimension $\D$ are also eigenstates of the Casimir, with eigenvalue $\D(\D-1)$. 

Since we want to find the specific values $\D_\star$ at which (\ref{eq:scalarpoles}) has a solution $\phi_\star$, we consider states with a well-defined $\D$.  Since these are Casimir eigenstates, we use $\D$ to label the representations we study. A particular state can be specified by its $L_0$ eigenvalue $\ell_0$ combined with $\D$. Note $L_\pm$ act as lowering/raising operators on the eigenvalue $\ell_0$. Additionally, since the quasinormal mode method will only work when the degeneracy of the states $\phi_\star$ is finite, we will insist that the representations have finite length.

If a representation labelled by $\D$ has finite length, then it must have a highest weight state satisfying
\begin{align}\label{eq:hwconditions}
L_0 \hstate &= h \hstate, 
\\ \nn
L_+ \hstate &= 0,
\end{align}
where $h$ here labels both the $L_0$ eigenvalue and the value of $\D$. Since we want a finite length representation, we also require $\left(L_-\right)^{p+1}\hstate =0, \, \left(L_-\right)^p \hstate \neq 0$, where $p+1$ is the length of the representation, and $p 
\in {\mathbb Z}_{\geq 0}$.  We can then use the commutator algebra to deduce
\begin{align}
[L_+,L_-^{p+1}]\hstate = L_+L_-^{p+1}\hstate &= (p+2h)(p+1)L_-^p\hstate
\\
\Rightarrow h&= -p/2\label{eq:useful1}.
\end{align}
In other words, $2h$ must be a non-positive integer, and the dimension of the representation with highest weight state $
\hstate$ is given by $p+1=2(-h)+1$.

We will also need expressions for the symmetry generators in specific coordinates.  
These are the Killing vectors of AdS$_2$.
We present these as vectors; the generators themselves are Lie derivatives acting in these directions. 
For the coordinates in Eq.~(\ref{eq:etathetametric}), we have
\begin{align}
L_0 &= \cos \theta \partial_\eta - \coth \eta \sin \theta \partial_\theta,
\label{L0Global}
\\
L_\pm &= i \sin\theta\partial_\eta +i \left(\coth\eta\cos\theta \mp 1 \right)\partial_\theta.
\label{LpmGlobal}
\end{align}
In Poincar\'e coordinates, with metric
\beq\label{eq:poincare}
ds^2 = \frac{dt^2+dz^2}{z^2},
\eeq they are
\begin{align}\label{eq:KillinginPoincare}
L_0 &= t\partial_t + z \partial_z,
\\ \label{eq:KillinginPoincare2}
L_- &= (t^2-z^2)\partial_t + 2 z t \partial_z,
\\\label{eq:KillinginPoincare3}
L_+ &= \partial_t.
\end{align}
The Killing vectors $L^\mu_i \partial_\mu$ (with $i=0,+,-$) act on a scalar through their Lie derivatives, i.e.
they act on a scalar function $\phi$ as $\calL_{L_i}\phi=L^\mu_i \partial_\mu \phi$.  For notational simplicity, we denote this action as $L_i \phi$.

\subsection{The scalar finite representations on AdS$_2$}
In order to study which values of $h$ and thus $\Delta_\star$ are actually exhibited in the scalar case, we use Poincar\'e coordinates.
A highest weight state must solve Eq. (\ref{eq:hwconditions}); for a scalar function $\phi_h(t,z)$ of weight $h$ these become
\begin{align}
{L_+} \phi_h &= \partial_t \phi_h = 0
\\
{L_0} \phi_h & = t\partial_t\phi_h + z\partial_z \phi_h = h \phi_h.
\end{align}
Solving these equations we find
\beq
\label{eq:scalarhighestweight}
\phi_h = z^h,
\eeq
where we have ignored overall normalization since it is irrelevant for our analysis.
Consider now 
\be
{L_-^{p+1}}\phi_h =0, \quad
{L_-^{p}}\phi_h \ne 0 \,,
\ee where $p+1$ is the length of the representation, and $p 
\in {\mathbb Z}_{\geq 0}$. Since ${L_-^{p}}\phi_h$ is an eigenfunction of ${L_0}$ with eigenvalue $h+p$, it must be of the form
\be
{L_-^{p}}\phi_h=\text{const} \times F(t/z) z^{h+p}
\ee for some function $F$. Solving ${L_-^{p+1}}\phi_h =0$ gives $F(x)=(1+x^2)^{h+p}$, which means
 \be
{L_-^{p}}\phi_h=\text{const} \times \left[1+\left(\frac{t}{z}\right)^2 \right]^{h+p}z^{h+p}\,.
\ee  The fact that $\phi_h$ is highest weight implies that ${L_+^{p+1}} 
{L_-^{p}}\phi_h \propto {L_+} \phi_h=0$. Remembering that $L_+^\mu\partial_\mu = \partial_t$, this condition reduces to
\be
\partial_t^{p+1}\left[1+\left(\frac{t}{z}\right)^2 \right]^{p/2}=0\,.
\ee 
where we have used the fact that $h=-p/2$ (with $2h$ a non-positive integer) from Eq.~(\ref{eq:useful1}).
If $p$ is a non-negative even integer, this equality is trivial since $(1+t^2)^{p/2}$ is a polynomial of degree $p$, and is thus annihilated by a $p+1$ derivative. If $p$ is instead a positive odd integer, then $(1+t^2)^{p/2}$ is some positive integer power of $\sqrt{1+t^2}$ and is not annihilated by any positive number of $t$-derivatives. Thus, $p$ must be a non-negative even integer, so $h=-p/2$ must be a non-positive integer. In short, we find that on AdS$_2$, the finite scalar representations consist of 
\be\label{eq:hstates}
\phi_h, \, 
L_- \phi_h,\, 
L_-^2 \phi_h,\,
\ldots , \, 
L_-^{-2h} \phi_h,
\ee
where $h \in {\mathbb Z}_{\leq 0}$.  Since these are the zero modes, with $\D_\star = h$ and degeneracy $D_h=-2h+1$, we have the same locations and degeneracies for the poles in Eq. (\ref{eq:weierstrass}).  Consequently, this algebraic method recovers the same answer for the partition function of the AdS$_2$ scalar as found previously in \cite{Keeler:2014hba}.

\subsection{Matching these scalar states to those from \cite{Keeler:2014hba}}
In the previous section, we did not impose boundary conditions on the functional form of the scalar states.  Instead, we simply insisted that the states in which we are interested should be in finite representations labelled by a fixed value of $\D$.  These restrictions resulted in the same number of states for each $\D$ as we found via boundary conditions in \cite{Keeler:2014hba}, shown here in Eq. (\ref{eq:explicitphistar}).

We now show that these two sets of states are related to each other via linear combination; consequently, the algebraic conditions do also impose the boundary conditions in Eq. (\ref{eq:bcandperiodic}).  The lowest case, when $h=0$, is actually quite trivial; both functions are just constants and thus equivalent up to overall normalization.

The next case, case $h=-1$, requires a bit more work.  In terms of the $\phi_{hl}$ defined in Eq. (\ref{eq:explicitphistar}), the highest weight state $\phi_{-1}$ becomes 
\beq
\phi_{-1} = \phi_{-1,-1} -2i  \phi_{-1,0} + \phi_{-1,1}.
\eeq
This can be checked two ways: first, via the (complicated) coordinate transformation between (\ref{eq:etathetametric}) and (\ref{eq:poincare}), and secondly by checking that the linear combination on the right hand side is a highest weight state, using the explicit expressions for $L_0,\, L_\pm$ in (\ref{L0Global}, \ref{LpmGlobal}).

More generally, the highest weight state is proportional to
\beq
\phi_h\propto\sum_{l=h}^{l=-h}\frac{(-h)!}{(-2i)^{|l|}|l|!(-h-|l|)!}\phi_{h,l}.
\eeq
Similarly, the descendants of these highest weight states can also be written as linear combinations of the $\phi_{h,l}$.  We can find the exact linear combination by noticing that $L_--L_+$ is an eigenoperator for $\phi_{h,l}$:
\beq\label{eq:lplusminus}
(L_--L_+)\phi_{h,l} = 2i \partial_\theta \phi_{h,l} = -2l \phi_{h,l}.
\eeq
It is additionally useful to recall that $L_+\phi_h=0$, and $L_0\phi_h=h\phi_h$.  Using these facts, we can write explicit expressions for $L_-^k\phi_h$ as linear combinations of the $\phi_{h,l}$, and each linear combination is unique; however the general expressions are not particularly illuminating so we do not reproduce them here.

Instead, we now move on to discuss the boundary conditions.  Since the highest weight states and their descendants can all be written as linear combinations of the $\phi_{h,l}$ from \cite{Keeler:2014hba}, they inherit their boundary conditions, namely smoothness at $\eta=0$, periodicity in $\theta$, and the falloff condition Eq. (\ref{eq:bcandperiodic}). In fact, the highest weight condition $L_+\phi_h=0$ together with the finite representation condition $L_-^{-2h+1}\phi_h=0$ impose both smooth regular behavior at the center of Euclidean AdS and the boundary condition at infinity.

We can use this fact to write the falloff condition Eq. (\ref{eq:bcandperiodic}) in a coordinate invariant manner.  States can always be labelled by the eigenvalues of a complete set of commuting operators; in AdS$_2$, we can choose the Casimir with eigenvalue $\Delta(\Delta-1)$ and $L_0$ with eigenvalue $\ell_0$.  The boundary conditions can then be rewritten as these eigenvalues satisfying
\be
|\ell_0|\leq |\Delta|.
\ee
If we choose instead $(L_--L_+)/2$ and the Casimir as the set of commuting operators, as in Eq (\ref{eq:lplusminus}), we similarly find $|l|\leq|h|$.  We are interested in representations such that all states in the representation obey the condition; this is equivalent to saying that the representations are of finite length.

\subsection{Generalization to higher dimensional AdS$_{2n}$}
\label{sec:higherscalar}
In \cite{Keeler:2014hba} for AdS$_{2n}$, the zero modes were obtained to be
\be
\label{eq:previouspaper1}
\Delta_\star=-p,~p=0,1,2,\ldots
\ee with degeneracy
\be\label{eq:previouspaper2}
D(p)=\frac{2p+d}{d} {p+d-1 \choose d-1} ,
\ee where $d+1=2n$. The algebraic method of finding finite dimensional representation of $SO(d+1,1)$, for $2n=d+1$, is most efficiently phrased in terms of finding the zero-eigenvalues of the inner-product matrix at each level. This calculation, though fairly straightforward, is extremely tedious at higher-levels. In Appendix \ref{sec:highershortscalar}, we have diagonalized the inner-product matrix for the case of AdS$_4$ and AdS$_6$ to find these finite representations for the first few levels. The results agree with Eq.~(\ref{eq:previouspaper1})-(\ref{eq:previouspaper2}). In this algebraic method, no explicit expressions of the zero modes are needed, in contrast to the original computations in Ref.~\cite{Keeler:2014hba}.


\section{Spin $\frac{1}{2}$ zero modes}
\label{sec:spinhalf}
In this section we compute the spin half finite representations in analogy with the the scalar case. We start with a spin half (Dirac spinor) highest weight state $|h\rangle$ and construct all of the states in the finite representations by repeated action with $L_-$. The action of the $SL(2,\mathbb{R})$ operators $L_0$, $L_{\pm}$ on spinors is achieved via Lie derivatives along the directions of those operators; i.e., if the vector $V = V^\mu \partial_\mu$ is an infinitesimal generator of the $SL(2,\mathbb{R})$ algebra, the Lie derivative along the direction of $V$ acting on a spinor is the infinitesimal representation of the $SL(2,\mathbb{R})$ algebra acting on the spinor representation.

\subsection{Lie derivatives and spinors}

The definition of a Lie derivative acting on a spinor along a Killing vector $V = V^\mu \partial_\mu$ is \cite{Choquet:2000}
\beq
\label{eq:liederivatives}
{\cal L}_V \psi = V^\mu \nabla_\mu\psi - \frac{1}{8}(\nabla_\mu V_\nu - \nabla_\nu V_\mu)\gamma^\mu \gamma^\nu \psi~.
\eeq
Highest weight states are eigenstates of ${\cal L}_{L_0}$ that are annihilated by ${\cal L}_{L_+}$, so the highest weight spinors $\psi$ must solve
\begin{align}
{\cal L}_{L_0} \psi &= h  \psi \,, \\
{\cal L}_{L_+} \psi&= 0.
\end{align}
We work in Poincar\'e coordinates as in Eq. (\ref{eq:poincare}) and choose the gamma matrices $\gamma^{\hat{a}} = \{\gamma^{\hat{t}} , \gamma^{\hat{z}}\}$, where hatted indices refer to frame indices, to be 
\begin{align}
\gamma^{\hat{t}} &= \sigma^1 = \left[ \begin{array}{ccc}
 0& 1 \\
 1 & 0 \end{array} \right], \\
 \gamma^{\hat{z}} &= -\sigma^2= \left[ \begin{array}{ccc}
 0& i \\
-i & 0 \end{array} \right].
\end{align}
The $SL(2,\mathbb{R})$ generators in Poincar\'e coordinates are given in Eq.~(\ref{eq:KillinginPoincare})-(\ref{eq:KillinginPoincare3}). The expression in Eq. (\ref{eq:liederivatives}) for the Lie derivative of a Killing vector $V^\mu\partial_\mu$ acting on a spinor $\psi$ can now be explicitly written as:
\begin{align}\label{eq:liedergenericglobal}
{\cal L}_V \psi &=\bigg( V^t (\partial_t - \frac{i\sigma^3}{2z}) + V^z \partial_z + \frac{i}{4}(\partial_t V_z - \partial_z V_t) z^2\sigma^3  \bigg) \psi.
\end{align}
At this point it is already clear why Poincar\'e coordinates are advantageous: for $V = L_+$, we have $V^t =1$, $V_t =\frac{1}{z^2}$ and $V^z =V_z=0$, such that the differential equation ${\cal L}_{L_+} \psi = 0,$ is simply
\be
\partial_t \psi = 0.
\ee
Moreover, the action of ${\cal L}_{L_i}$ on a two component spinor $\psi$ with upper component $\phi_1$ and lower component $\phi_2$ is
\beq
L_i\psi = \calL_{L_i}\begin{bmatrix}\phi_1 \\ \phi_2\end{bmatrix}.
\eeq
In Poincar\'e coordinates, the action of the $SL(2,\mathbb{R})$ generators $L_0, L_\pm$ on $\psi$ becomes
\begin{align} \label{eq:liederonspinors}
{L_0} \psi = \begin{bmatrix}{L_0}\phi_1 \\ {L_0}\phi_2\end{bmatrix},
~{L_+} \psi  =\begin{bmatrix} {L_+}\phi_1 \\ {L_+}\phi_2\end{bmatrix},
~{L_{-}} \psi =\begin{bmatrix}({L_{-}}+iz)\phi_1 \\ ({L_{-}}-iz)\phi_2\end{bmatrix}.
\end{align}
Since $L_0$ and ${L_+}$ act independently on the top and bottom components of the spinor, the problem of finding spin half highest weight states is quite similar to the problem of finding scalar highest weight states.

As one acts repeatedly with ${L_{-}}$ to find all states in a finite representation there is a departure from the scalar case due to the extra terms in the expression of ${\cal L}_{L_{-}} \psi$. We will see that this departure manifests itself mainly in the number of states in the finite representations, which is to be expected for a representation with a different spin. 

\subsection{Finite representations}
\label{sec:spinorads2}
We consider a highest weight state $\psi_h$ such that 
\begin{align} \nn
{L_0} \psi_h &=h \psi_h, \\ \nn
{L_+} \psi_h &=0. 
\end{align}
The condition ${L_+} \psi_h =0$ requires both components of $\psi_h$ to be independent of time, 
\be
{L_+} \psi_h = \begin{bmatrix} \partial_t \phi_1 \\ \partial_t \phi_2\end{bmatrix} =0.
\ee
The condition ${L_0} \psi_h =h \psi_h$ requires $\psi_h$ to be of the form
\be
\psi_h = z^h \begin{bmatrix}  c_1 \\ c_2\end{bmatrix},
\ee
where $c_1, c_2$ are constants. As shown in Eq. (\ref{eq:useful1}), these conditions together with the commutation relations require $h = - p/2$, for $p \in {\mathbb Z}_{\geq 0}$.  We find the same result since by definition the Lie derivatives satisfy the same commutation relations; however, we now show that the functional form of the spinor  modes imposes further restrictions on the values that h can take. 

The state ${L_-}^p \psi_h$ is an eigenstate of ${L_0}$ with eigenvalue $h+p$, so ${L_-}^p \psi_h$ must be of the form
\be
{L_-}^p \psi_h= z^{h+p} \begin{bmatrix}   c_1G(t/z)  \\ c_2G^*(t/z)\end{bmatrix},
\ee
where $G(t/z), G^*(t/z)$ are functions of the combination $t/z$. We can solve the condition ${L_-}^{p+1} \psi_h = 0$ to find
\be
G(t/z) =[1+(t/z)^2]^{\frac{p-1}{2}} [ 1+ i(t/z) ],
\ee
and $G^*(t/z)$ must be its complex conjugate.

The state ${L_+}^{p+1}{L_-}^p \psi_h$ vanishes, since ${L_+} \psi_h=0$. Thus, we find
\be
 (\partial_t)^{p+1} \left( [1\pm i(t/z)][1+(t/z)^2]^{\frac{p-1}{2}} \right)=0.
\ee
For clarity let us analyze the real and imaginary parts separately:
\begin{align} \label{eq:realpart}
(\partial_t)^{p+1} \left([1+(t/z)^2]^{\frac{p-1}{2}} \right) &=0, \\ \label{eq:imaginarypart}
(\partial_t)^{p+1} \left((t/z) [1+(t/z)^2]^{\frac{p-1}{2}}\right) &=0.
\end{align}
In order to be killed by $p+1$ derivatives, the function of $t$ must to be a polynomial of degree $p$ or lower. If $p$ is even, $[1+(t/z)^2]^{\frac{p-1}{2}}$ is not polynomial and in fact is not killed by $(\partial_t)^{p+1}$. Hence $p$ must be an odd number, in which case $[1+(t/z)^2]^{\frac{p-1}{2}}$ and $(t/z) [1+(t/z)^2]^{\frac{p-1}{2}}$ are polynomial with degrees $p-1$ and $p$ respectively. 

We already knew from the commutation relations that $2h$ must be a nonpositive integer; we have now shown that the functional form of the spinor modes additionally requires that $2h$ is odd, or $h= -p - 1/2$ for for $p \in {\mathbb Z}_{\geq 0}$.

If we consider each chirality separately, then for a given $p$, we have $2(-h)+1=2(p+1)$ states:
\be \label{eq:spinhalfmodes}
\psi_h,\quad{L_{-}} \psi_h,\quad {L_{-}}^2 \psi_h, \quad \ldots, \quad{L_{-}}^{2p+1} \psi_h,
\ee 
with ${L_0}$ eigenvalues ranging from $h=-1/2-p$ to $-h=1/2+p$. The states with ${L_0}$ eigenvalues greater than $-h$ are annihilated in analogy with the scalar case; the principal difference with respect to the spin zero representations is the number of states.

\subsection{Sum over modes}

In this section, we consider the partition function for a Dirac fermion. We take the spin half modes found previously (\ref{eq:spinhalfmodes}) and sum over them by adapting the formula (\ref{eq:logZ}). In the derivation of (\ref{eq:logZ}) we remarked that the partition function had only poles (see  Eq.~(\ref{eq:partifunctionscalar})). This time we are computing a fermionic determinant,
\beq\label{eq:Zfermion}
Z \propto \det [\slashed{\nabla} - m ],
\eeq
so the modes we computed correspond to zeros in the spin half partition function, and there are no poles. Here $m$ is again a function of $\D$, given by $m=\Delta-\half$.

Following the logic we used for the scalar field, we use Weierstrass's factorization theorem to write
\be \label{eq:weierstrasshalf}
Z(\Delta) = e^{\text{Pol}(\D)}\prod_{h}(\D - h)^{D_{h}}.
\ee
The states we found are at $-h=-\Delta_\star = p +\frac{1}{2}$, $p=0,1,2,\ldots$, with degeneracies $4(p+1)$, where $2(p+1)$ modes are coming from each chirality. However, only one chirality should be accounted to match the spinor representations of the conformal group \cite{Henneaux:1998ch,Henningson:1998cd,Iqbal:2009fd}.
 We insert those values in (\ref{eq:weierstrasshalf}) and take the log,
\begin{align} \label{eq:logzfermion}
\log{Z} &= \textrm{Pol}(\D) + \sum_{p=0}^\infty(2p+2)\log\bigg(\D+p+\frac{1}{2}\bigg), \\ \nn
& = \textrm{Pol}(\D) - 2\zeta^\prime(-1,\D + \frac{1}{2}) + (2\D -1)\zeta^\prime(0,\D+\frac{1}{2}),
\end{align} 
where we have again used the Hurwitz zeta function to regularize the sum. We now proceed to find the asymptotics of $\log Z$ and evaluate $\textrm{Pol}(\D)$. First we rewrite $\log Z $ in terms of the mass $m = \D -\frac{1}{2}$ and expand around large $m$,
\beq\label{eq:largemassdhs}
\log Z = \textrm{Pol}(m) - \frac{3}{2}m^2 + \frac{1}{2}m^2\log{m^2} - \frac{1}{12}\log(m^2) - \frac{1}{120m^2} - \mathcal{O}(m^{-5}).
\eeq
To compute $\textrm{Pol}(m)$ we match our expression for large $m$ with the heat kernel curvature expansion of a free spin-half field in AdS$_2$ \cite{Vassilevich:2003xt},
\begin{align}
\log Z &= -\frac{1}{4\pi} \int_{\mathbb{H}^2}\sqrt{g}\bigg( \int_{e^{-\gamma}\Lambda^{-2}}^\infty \frac{dt}{t^2}e^{-tm^2} - \frac{R}{12} \int_{e^{-\gamma}\Lambda^{-2}}^\infty \frac{dt}{t}e^{-tm^2}\bigg) + \mathcal{O}(m^{-1}) \\ \nn
 &= - \frac{1}{4\pi} \int_{\mathbb{H}^2}\sqrt{g}\bigg( e^\gamma \Lambda^2 - m^2 + m^2\log(\frac{m^2}{\Lambda^2}) + \frac{R}{12}\log(\frac{m^2}{\Lambda^2})\bigg) + \mathcal{O}(m^{-1}) + \mathcal{O}(m/\Lambda)
\end{align}
The relevant Seeley-DeWitt coefficients are $a_0 =1$ and $a_2 = -\frac{R}{12}$. We introduced the cutoff $e^{-\gamma}\Lambda^{-2}$, where $\Lambda$ is a quantity with dimensions of mass, and $\gamma$ is the Euler-Mascheroni constant. The overall minus sign is due to the fact that we are computing a fermionic determinant.

The Ricci scalar of AdS$_2$ with unit radius is $R=-2$ and the regularized AdS$_2$ volume is $-2\pi$. We drop the $m$-independent term and insert the values for $R$ and the volume of AdS$_2$ into the heat kernel expansion, obtaining 
\begin{align}\label{eq:finalheatkernelspinor}
\log Z &= -\frac{1}{2}m^2+\frac{1}{2}m^2\log\bigg(\frac{m^2}{\Lambda^2}\bigg) -\frac{1}{12}\log\bigg(\frac{m^2}{\Lambda^2}\bigg) + \mathcal{O}(m^{-1}) + \mathcal{O}(m/\Lambda).
\end{align}
We match expressions (\ref{eq:largemassdhs}) and (\ref{eq:finalheatkernelspinor}) and find $\textrm{Pol}(m)$ ,
\begin{align}\label{eq:Polspinor}
 \textrm{Pol}(m) &=m^2-\frac{1}{2}m^2\log(\Lambda^2) +\frac{1}{12}\log(\Lambda^2), \\ \nn
 &= \bigg(\D-\frac{1}{2}\bigg)^2-\frac{1}{2}\bigg(\D-\frac{1}{2}\bigg)^2\log(\Lambda^2) +\frac{1}{12}\log(\Lambda^2).
\end{align}
In the last step we rewrote Pol in terms of the conformal dimension $\D$. Now that we found Pol, we insert its expression in the formula for the partition function (\ref{eq:logzfermion}) completing our computation,
\beq \label{eq:spinhalfresult}
\log{Z}  =\bigg(\D-\frac{1}{2}\bigg)^2-\frac{1}{2}\bigg(\D-\frac{1}{2}\bigg)^2\log(\Lambda^2) +\frac{1}{12}\log(\Lambda^2) - 2\zeta^\prime(-1,\D + \frac{1}{2}) + (2\D -1)\zeta^\prime(0,\D+\frac{1}{2}).
\eeq
This is the partition function of a free Dirac fermion in AdS$_2$.

We close this section providing a explicit check with results previously computed by other methods. In \cite{Banerjee:2010qc}, Banerjee, Gupta, and Sen compute the heat kernel density for a free Dirac fermion on AdS$_2$; their result is
\beq \label{eq:senkernel}
K(t) = -\frac{1}{2\pi t}\bigg( 1 +\frac{1}{6}t-\frac{1}{60}t^2 + \mathcal{O}(t^3)\bigg).
\eeq
We integrate each term of the heat kernel (\ref{eq:senkernel}) after inserting the mass factor $e^{-tm^2}$,
\beq
\log Z = \frac{{\rm{Vol}_{\rm AdS_2}}}{2}\int_\epsilon^\infty \frac{dt}{t}K(t)e^{-tm^2},
\eeq
where the factor of the regularized volume of AdS$_2$ arises because Ref.~\cite{Banerjee:2010qc} computes a heat kernel density $K(t)$.
Expanding around small $\epsilon$, the result is 
\begin{align}
\log Z   &=  -\frac{ {\rm{Vol}_{\rm AdS_2}}}{4\pi }\bigg(  \frac{1}{\epsilon} - m^2 + m^2 \gamma+ \frac{1}{6} (-\gamma - \log(m^2 \epsilon)) + m^2 \log(m^2 \epsilon) -\frac{1}{60 m^2} + \mathcal{O}(m^{-4}) \bigg)  \\ 
&=   - \frac{1}{2}m^2 - \frac{1}{12} \log\bigg(\frac{m^2}{\Lambda^2}\bigg) + \frac{1}{2}m^2 \log\bigg(\frac{m^2}{\Lambda^2}\bigg)-\frac{1}{120 m^2} + \mathcal{O}(m^{-4}) + \mathcal{O}(\Lambda^{-2}). \label{eq:senresult}
\end{align}
In the second step we have set $\epsilon = e^{-\gamma}\Lambda^{-2}$ as in our computation.  Comparison of (\ref{eq:senresult}) with (\ref{eq:largemassdhs}) shows that the logarithmic terms agree. Moreover, insertion of the polynomial terms (\ref{eq:Polspinor}) in (\ref{eq:largemassdhs}) yields agreement between the polynomial terms computed by \cite{Banerjee:2010qc} as well. 

In conclusion, the partition function (\ref{eq:spinhalfresult}) we computed agrees in the large mass limit with the previous results of \cite{Banerjee:2010qc}.

\subsection{Generalization to AdS$_{4}$}
Similarly to the scalar case (as discussed in \ref{sec:higherscalar}), the algebraic method of finding finite dimensional spinor representations of $SO(d+1,1)$ involves finding the zero-eigenvalues of the inner-product matrix at each level for a spinor highest weight representation. We have diagonalized the inner-product matrix for the case of AdS$_4$ to find these finite representations up to a few levels. 
The details are provided in Appendix~\ref{sec:higherspinor}.
The result can be summarized as
\be
h=-\half -p,\quad p=0,1,2\ldots
\ee with degeneracy for each $p$ given by\footnote{Furthermore, just as for the computations of the AdS$_2$ spinor, we have found the zero mode spinor eigenfunctions satisfying appropriate boundary conditions and checked that, indeed, we reproduced  the AdS$_4$ spinor finite representations with the correct degeneracy.}
\be
D(p)=\frac{2}{3} (p+1) (p+2) (p+3).
\ee
In fact, we can use these results alone to recover the logarithmic portion of the spinor one loop effective action in the existing literature.
We first use Eq.~(\ref{eq:logZ}) with $-h=-\D_\star = p +\frac{1}{2}$, $p=0,1,2,\ldots$ and degeneracies $D(p)=\frac{2}{3} (p+1) (p+2) (p+3)$ to find
\bea
\log{Z} &=& \textrm{Pol}(\D) +\frac{2}{3} \sum_{p=0}^\infty (p+1) (p+2) (p+3)\log\bigg(\D+p+\frac{1}{2}\bigg), \\
& =& \textrm{Pol}(\D) 
-\frac{2}{3}\left[
\zeta'\left(-3,\D+\half\right)
+3\left(2-\D\right)\zeta'\left(-2,\D+\half\right)
\right.
\\ 
&&\left.
+\left(3\D^2-12 \Delta+11\right) \zeta'\left(-1,\D+\half\right)
- (\D -1)(\D -2)(\D -3) 
\zeta'\left(0,\D+\half\right)
\right]\,.\nonumber
\eea
In the large mass expansion, using $\D=2+m $ and expanding around large $m$ gives
\beq\label{eq:largemassdhsads4}
\log Z = \textrm{Pol}(m) -\frac{25 }{72}m^4+\frac{1}{2}m^2+\frac{1}{24}\left(2m^4-4m^2+\frac{11}{15}\right) \log (m^2)+ \mathcal{O}(m^{-2})\,,
\eeq which reproduces the $\log m^2$ terms in Ref.~\cite{PhysRevD.45.3591} upon regularizing the volume of AdS$_4$ to be $(4 \pi^2)/3$.


\section{Massive spin-one and spin-two finite representations}
\label{sec:spinoneandtwo}
In this section, we will show that finite representations for spin-one and spin-two fields on AdS$_2$ are directly related to the scalar finite representations on AdS$_2$.  

The spin one and spin two fields we consider are massive, since the quasinormal mode method relies on treating the partition function as a function of a complex mass parameter.  Accordingly, the fields we consider have no gauge symmetry; in order to apply these results to the massless case, we would need to perform gauge-fixing and separately treat the contributions of the associated ghosts on their own.

To start, let us take $\phi_h$ to be a scalar highest weight mode, i.e. $L_0 \phi_h= h \phi_h$ and $L_+\phi_h=0$, with representation length $D_h$. We will consider spin one and spin two separately in the following subsections.

\subsection{Spin one}
 Let $\fA\equiv A_\mu dx^\mu$ be a one-form. In differential form notation, for an arbitrary vector $\xi^\mu \partial_\mu$, the Lie derivative acts on $\phi_h$ and a one-form $\fA$ by the usual rule:
\be
\calL_\xi \phi_h = i_\xi d\phi_h ,\quad
\calL_\xi \fA = d(i_\xi \fA) + i_\xi d\fA, 
\ee where $i_\xi$ is the interior product/contraction.

Next, define the one-form
\be
\fA_h \equiv d\phi_h \,.
\ee Then
\be
\calL_{L_0} \fA_h=d[\calL_{L_0} \phi_h] = h \fA_h,\quad
\calL_{L_+} \fA_h = d[\calL_{L_+} \phi_h]=0\,,
\ee i.e. $\fA_h$ is an highest weight spin one field.
Furthermore, since
\be
\calL_{L_-^k} \fA_h=d(\calL_{L_-^k} \phi_h)
\ee if $\phi_h$ is the scalar highest weight state for a finite representation with dimension $D_h$, then $\fA_h$ is a spin-one highest weight state for a finite representation with the same dimension $D_h$.

On the other hand, consider
\be
(\tilde{A}_h)_\mu \equiv \epsilon_{\mu\nu} \nabla^\nu \phi_h\, ,
\ee then since
\be
\calL_\xi (\tilde{A}_h)_\mu = \epsilon_{\mu}{}^\nu \calL_\xi (A_h)_\nu
\ee we have that
\be
\calL_{L_0}(\tilde{A}_h)_\mu=h (\tilde{A}_h)_\mu,\quad
\calL_{L_+} (\tilde{A}_h)_\mu = 0,\quad
\ee as well as
\be
\calL_{L_-^k} (\tilde{A}_h)_\mu = \epsilon_{\mu}{}^\nu \calL_{L_-^k} (A_h)_\nu=\epsilon_{\mu}{}^\nu d\left( \calL_{L_-^k} \phi_h \right),
\ee which implies that  if $\phi_h$ is the scalar highest weight state of a finite representation with dimension $D_h$, then $(\tilde{A}_h)_\mu$ is a spin-one highest weight state of a finite representation with the same dimension $D_h$.

 So far, we have exhibited two set of modes for each $h$, i.e. $A_h$ and $(\tilde{A}_h)$. If they are independent modes, then
due to the fact that the highest weight conditions and the finite representation conditions are two-component (note that we are in AdS$_2$) first-order differential equations, we have obtained the most general solutions by taking linear combinations of these two independent solutions
$c_1 (A_h)_\mu +c_2 (\tilde{A}_h)_\mu $. Indeed, for $h\ne 0$, $A_h$ and $(\tilde{A}_h)$ are independent. For $h=0$, however, recall from Eq.~(\ref{eq:scalarhighestweight}) that $\phi_h=$ constant, and hence $A_h=\tilde{A}_h=0$, which means that these are not the non-trivial highest weight modes that we are after. Furthermore, for $h=0$, one can explicitly use the AdS$_2$ Killing vectors to show that it is impossible to have a finite highest weight representation.

Thus the zero modes of a massive spin one field on AdS$_2$ are the same as that of the scalar, but we have {\it twice} the degeneracy together with the restriction that $h\ne 0$. 
For massive spin one fields in two dimensions, the relation between conformal dimension $\Delta$ and mass $m$ is as in the scalar case: $m^2=\Delta(\Delta-1)$. Note that the non-existence of $h=0$ zero modes implies that the one-loop determinant of a massive spin-one is twice that of the scalar one, up to an extra term which corrects for the fact that $h=0$ zero modes are not present in the massive spin-one case. Explicitly, this extra term gives exactly a $\log \Delta$ contribution in the one-loop determinant of a {\it real} massive vector field. 

On the other hand, from Ref.~\cite{Banerjee:2010qc}, in computing the logarithm of partition function for a real massless spin-one field using the heat kernel method, one observes that there is a subtlety coming from extra square-integrable ``zero-eigenvalue" modes. This effect accounts for the difference between the vector heat kernel and twice the scalar heat kernel. In fact, from the heat kernel of the massless case, one can easily compute the {\it massive} spin-one one-loop determinant.
The correction from the zero-eigenvalues modes translates into a $\log \Delta$ term which is exactly the same as the contribution $h=0$ zero modes we mentioned in the previous paragraph.\footnote{We thank the referee for pointing out this subtle effect which enabled us to do a more careful analysis on the non-existence of the $h=0$ spin-one zero mode.}
 
It is interesting that although the two methods reproduce the same subtle corrections, the origin of this term in our method has to do with removing some modes from the scalar case (much like the $S^2$ zero-eigenvalue mode's removal which is discussed around Eq.~3.1.10 of Ref.~\cite{Banerjee:2010qc}), whereas in Ref.~\cite{Banerjee:2010qc} this term comes from additional square integrable zero-eigenvalues.

\subsection{Spin two}
Similarly to the strategy in the spin one case, since a symmetric two-tensor $h_{\mu\nu}$ has 3 independent components in AdS$_2$ and the highest weight conditions are first order, it is sufficient to show that there are 3 independent solutions, each of which is in one-to-one correspondence with the scalar highest weight.

First, consider $h\ne0$ and let $A_h$ as well as $(\tilde{A}_h)$ be the two highest-weight spin-one fields considered in the previous subsection. Then the two spin-two highest-weight modes can be obtained from the following form of $h_{\mu\nu}$:
\be
h_{\mu\nu} = \calL_{A_h} g_{\mu\nu}\quad
\mbox{or}\quad
h_{\mu\nu} = \calL_{\tilde{A}_h} g_{\mu\nu}\quad
\ee 
where $g_{\mu\nu}$ is the AdS$_2$ metric. A short computation shows that
\bea
\calL_{L_0} h_{\mu\nu} &=&\calL_{[L_0,A_h]} g_{\mu\nu} =h\calL_{A_h} g_{\mu\nu} =h h_{\mu\nu}\nonumber\\
\calL_{L_+} h_{\mu\nu} &=&\calL_{[L_+,A_h]} g_{\mu\nu}=0\nonumber\\
\calL_{L_-^k} h_{\mu\nu} &=&\calL_{[L_-^k,A_h]} g_{\mu\nu},
\eea 
where we have used $\calL_{L_0} g_{\mu\nu}=0$ and the fact that $\calL_{L_0} \fA_h= h\fA_h$. Similar results hold for $\tilde{A}_h$. This shows that these highest weight spin-two modes are in one-to-one correspondence with the spin one highest weight (which in turn is in bijection with two copies of the scalar highest weight).  Furthermore, we can explicitly check that they are two independent spin-one highest weight modes, so we get two independent spin-two highest weight modes here, each with degeneracy $D_h$. 

Finally the third highest-weight mode comes from considering
\be
h_{\mu\nu} = \phi_h g_{\mu\nu} \,.
\ee By Leibniz's rule and the fact that $\calL_{L_0} g_{\mu\nu}=\calL_{L_+} g_{\mu\nu}=\calL_{L_-} g_{\mu\nu}=0$, we obtain
\bea
\calL_{L_0} h_{\mu\nu} &=& g_{\mu\nu} \calL_{L_0} \phi_h = h h_{\mu\nu}, \nonumber\\
\calL_{L_+} h_{\mu\nu} &=& g_{\mu\nu} \calL_{L_+} \phi_h = 0, \nonumber\\
\calL_{L_-^k} h_{\mu\nu} &=& g_{\mu\nu} \calL_{L_-^k} \phi_h \,.
\eea Thus, this spin-two highest weight solution is in one-to-one correspondence with the scalar modes.

 For $h=0$, the mode proportional to the metric (i.e. $\phi_h g_{\mu\nu}$) still exists while the modes  $\calL_{A_h} g_{\mu\nu}$ and $\calL_{\tilde{A}_h} g_{\mu\nu}$ vanish and thus they are not non-trivial zero modes. One can in fact  demonstrate that for $h=0$ the only finite-dimensional representation is given by the $\phi_h g_{\mu\nu}$ mode using explicitly the highest-weight equations on AdS$_2$.

In short, for $h\ne 0$ we have exhibited three independent highest-weight spin-two modes, and they all come from scalar highest-weight modes. For $h=0$, however, we only have one highest-weight mode. Similar to the spin-one case in the previous section, this implies that the log of partition function of a massive spin-two is given by three times the scalar one up to a log mass-squared correction. It would be interesting to compare this to the difference between the spin-two heat kernel and the scalar one, which possibly originates from zero-eigenvalue square-integrable modes, similar to the phenomenon observed for the spin-one case in Ref.~\cite{Banerjee:2010qc}.



\section*{Acknowledgements}
We would like to thank Gino Knodel and Eric Perlmutter for discussions during the conception of this project. C. K. is supported by the European Union's Horizon 2020 research and innovation programme under the Marie Sk\l{}odowska-Curie grant agreement No 656900.
P. L. is supported in part by the US Department of Energy under grant DE-SC0007859.
G. N. was supported by an NSERC Discovery Grant.

\appendix

\section{Finite representations of $SO(d+1,1)$}
\label{sec:app1}
Let us study a Euclidean CFT$_d$ on $R^d$ with coordinates $x^\mu$ for $d=2n-1$.
The $d$-dimensional Euclidean conformal group is generated by the dilatation $D$, translations $P_\mu$, special conformal transformations $K_\mu$ and the $SO(d)$-rotation subalgebra $M_{\mu\nu}$.\footnote{For $d=1$, we have that $D=L_0$, $K=L_+$ and $P=L_-$.} They satisfy the algebra
\bea\label{eq:confalgebra}
[D,P_\mu]&=& P_\mu,\quad [D,K_\mu]= -K_\mu,\quad
[K_\mu , P_\nu ] =2 \left( \delta_{\mu\nu} D - i M_{\mu\nu}\right) ,\nonumber\\
~[M_{\mu\nu}, P_\rho]& =& i (\delta_{\mu\rho} P_\nu - \delta_{\nu\rho} P_\mu),\quad
[M_{\mu\nu}, K_\rho] = i (\delta_{\mu\rho} K_\nu - \delta_{\nu\rho} K_\mu), \nonumber\\
~[M_{\mu\nu}, M_{\rho\sigma}] &= &i (\delta_{\mu\rho}M_{\nu\sigma}+\delta_{\nu\sigma} M_{\mu\rho} - \delta_{\mu\sigma} M_{\nu\rho}-\delta_{\nu\rho} M_{\mu\sigma}), 
\eea with the rest of the commutators being zero. In radial quantization,  the hermitian conjugate $\dag$ acts as
\be
M_{\mu\nu} = M^\dag_{\mu\nu},\quad P_\mu = K_\mu^\dag,\quad D^\dag=D.
\ee 
Using this algebra, we will study finite representations for a scalar highest weight state as well as a (Dirac) spinor highest weight state.

\subsection{Scalar}
\label{sec:highershortscalar}
Consider the highest weight representations of the conformal group with the highest weight scalar state $\hstate$ satisfying
\be
D \hstate = h \hstate,\quad M_{\mu\nu} \hstate =K_\mu \hstate=0.
\ee Descendants of the form $P_{\mu_1}\ldots P_{\mu_k} \hstate$ generate a complete set of states with $D=h+k$. We call these state level $k$ descendants of $\hstate$. At generic values of $h$, no finite-dimensional representations exist. However, at special (non-unitary) values of $h$, one might encounter a highest weight state (or null state). This means that we should quotient out (or set to zero within this representation) those null states and their descendants, which results in a finite representation. This representation is called a short/finite representation. We aim to study these representations. 

To do so, one turns to the computation of the inner-product matrix at each level.  As an illustrative example, we shall first first work out the case for $d=1$ and then go on to the case of $d =3$ and $d=5$.
\subsubsection{$d=1$}
\label{eq:scalard1}
For $d=1$, we only have one raising operator $P\equiv P_1$ and one lowering operator $K\equiv K_1$ and there are no $M_{\mu\nu}$'s.
The inner-product at level $k$ is given by
\be
M(k)\equiv \hbra K^k P^k \hstate=\left[2(h+k-1)+2(h+k-2)+\ldots+2h\right]M(k-1) =\sum_{p=0}^{k-1} 2(h+p)M(k-1) \,,
\ee implying that
\be
M(k)=\frac{\Gamma (k+1) \Gamma (2 h+k)}{\Gamma (2 h)}\,.
\ee We see that for $h=-p/2,\quad p=0,1,2,\ldots$, the states up to and including level $p$ have non-zero norm whereas $M(k)=0$ for $k>p$. Thus, we have a finite representation whenever $h=-p/2$ with dimension $(p+1)=2(-h)+1$ .

\subsubsection{$d=3$ and $d = 5$}
For $d>1$, this inner-product is not diagonal, so we have to diagonalize the inner-product matrix at level $k$
\be
M(k)_{\mu_1\ldots \mu_k, \nu_1\ldots \nu_k}\equiv 
\hbra K_{\mu_1} K_{\mu_2}\ldots K_{\mu_k} P_{\nu_k} \ldots P_{\nu_2}P_{\nu_1} \hstate \,.
\ee At low levels, one can calculate this rather straightforwardly.  For example
\bea
M(1)_{\mu_1,\nu_1}&\equiv &
\hbra K_{\mu_1} P_{\nu_1} \hstate =2h \delta_{\mu_1 \nu_1} \\\nonumber
M(2)_{\mu_1\mu_2,\nu_1\nu_2}&\equiv &
\hbra K_{\mu_1}K_{\mu_2} P_{\nu_2} P_{\nu_1} \hstate =4h(h+1) (\delta_{\mu_2 \nu_2}\delta_{\mu_1 \nu_1}+\delta_{\mu_2 \nu_1} \delta_{\nu_1 \mu_1})-4h \delta_{\mu_1 \mu_2} \delta_{\nu_1 \nu_2}
\eea where we have normalized $\langle h|h\rangle=1$. Inner-product matrices of higher level can be obtained straightforwardly with longer expressions in terms of products of delta functions. The explicit expressions are long and not enlightening. We diagonalize them using Mathematica for dimension $d=3$ and $d=5$ with levels up to level $4$, and get the following eigenvalues:\footnote{We are not displaying the zero eigenvalues due to antisymmetric states (for example, at level $2$,  $(P_i P_j- P_j P_i)\hstate=0$) since they are irrelevant. Furthermore, we have dropped any overall $h$-independent prefactor.
}
\begin{itemize}
\item For $d=3$:
\be
\begin{array}{|c | l |c |}
\hline
k& \text{Eigenvalues} & \text{Multiplicity}\\
\hline
  0 & 1 & 1 \\
\hline
  1 & h & 3 \\
\hline
  2 & h (  h-\half) & 1 \\
\hline
   &h (h+1) & 5 \\
\hline
  3 & h ( h+1) ( h-\half)  & 3 \\
  \hline
   & h (h+1) ( h+2) & 7 \\
  \hline
  4 &h (h+1) (h-\half) (h+\half)& 1 \\
  \hline
   &h  (h+1) ( h+2) (h-\half) & 5 \\
  \hline
   &h (h+1) (h+2) ( h+3)& 9 \\
     \hline
\end{array}
\ee

\item For $d=5$:
\be
\begin{array}{|c | l |c |}
\hline
k& \text{Eigenvalues} & \text{Multiplicity}\\
\hline
  0 & 1 & 1 \\
\hline
  1 & h & 5 \\
\hline
  2 &h ( h-\frac{3}{2}) & 1 \\
\hline
   &h (h+1) & 14 \\
\hline
  3 & h ( h+1) ( h-\frac{3}{2})  & 5 \\
  \hline
   & h (h+1) (h+2) & 30 \\
  \hline
  4 &h (h+1) (h-\half) (h+\half)& 1 \\
  \hline
   &h  (h+1) ( h+2) (h-\half) & 5 \\
  \hline
   &h (h+1) (h+2) ( h+3)& 9 \\
  \hline
\end{array}
\ee
\end{itemize}
With these data, we observe that for $h$ being a non-positive integer, the representation is shortened to be finite dimensional.
Here are the list of such non-positive $h$'s and their dimensions:
\begin{itemize}

\item For $d=3$:
\be
\begin{array}{|c | c|}
\hline
-h& \text{Dimension} \\
\hline
  0 & 1 \\
    \hline
  1 & 1+3+1=5 \\
    \hline
  2 & 1+3+6+3+1=14 \\
    \hline
    \end{array}
\ee
The pattern seems to be that for $h=-p$, the dimension is
\bea
&&1+3+6+10+\ldots+(p+1)(p+2)/2+\ldots+10+6+3+1\nonumber\\
&=&\frac{(p+1)(p+2)}{2}+2\sum^p_{q=0} (q+1)(q+2)/2  =\frac{(p+1)(p+2)(2p+3)}{6} \nonumber\\
&=&\left.\frac{2p+d}{d} {p+d-1 \choose d-1} \right|_{d=3}\,.
\eea

\item For $d=5$:
\be
\begin{array}{|c | c|}
\hline
-h& \text{Dimension} \\
\hline
  0 & 1 \\
    \hline
  1 & 1+5+1=7 \\
    \hline
  2 & 1+5+15+5+1=27\\
    \hline
    \end{array}
\ee
The pattern seems to be that for $h=-p$, the dimension is
\bea
\left.\frac{2p+d}{d} {p+d-1 \choose d-1} \right|_{d=5}\,.
\eea

\end{itemize}
In summary, we have obtained some evidence suggesting that for a general AdS$_{d+1}$, the finite representation with a scalar highest weight $h$ is given by $h=-p,~p=0,1,2,\ldots$ with dimension of the representation given by
\be
\frac{2p+d}{d} {p+d-1 \choose d-1} \,.
\ee  These values of $h$ (and the dimension of their representations) coincide with the zero modes obtained in \cite{Keeler:2014hba}.
It will be interesting to relate the results in this section (or the zero-modes) to the rational representations of conformal blocks \cite{Kos:2013tga,Penedones:2015aga,Iliesiu:2015akf}.

\subsection{Spinor}
\label{sec:higherspinor}
Consider the highest weight representation of the conformal group with the highest weight spinor state $\hstatespin$ (with $a$ being an index in the spinor representation of the $SO(2n)$) satisfying
\be
D \hstatespin = h \hstatespin,\quad K_\mu \hstatespin=0,\quad
M_{\mu\nu} \hstatespin =
-\sum_{b=1}^{2^{n}} (\Sigma_{\mu\nu})_{ab} \hstatespinb ,\quad 
\langle h,a | h,b \rangle
=\delta_{ab}\,,
\ee  where 
\be
\Sigma_{\mu\nu} = -\frac{i}{4}\left[
\gamma_\mu ,\gamma_\nu
\right]
\ee and the $\gamma_\mu$'s form representation of the Clifford algebra
\be
\left\{ \gamma_\mu ,\gamma_\nu \right\}=2\delta_{\mu\nu}\,.
\ee We follow the conventions in Chapter 3.1 of \cite{freedman2012supergravity} for Euclidean $\gamma_\mu$. The convention for $\Sigma_{\mu\nu}$ is such that it satisfies the same algebra as  $M_{\mu\nu}$ in Eq.~(\ref{eq:confalgebra}). 

In particular, for $d=3$ (or $n=1$), we have 2-component spinors and $2\times 2$ gamma matrices chosen as
\bea
\gamma^1 &=& \sigma_1= \left( \begin{array}{ccc}
 0& 1 \\
 1 & 0 \end{array} \right)\nonumber\\
\gamma^2 &=& \sigma_2= \left( \begin{array}{ccc}
 0& -i \\
i & 0 \end{array} \right) \nonumber\\
\gamma^3 &=& \sigma_3=  \left( \begin{array}{ccc}
 1& 0 \\
 0 & -1 \end{array} \right) \,,
\eea where $\sigma_i$'s are the Pauli matrices.
The rest of the structures (e.g. how to build descendants and etc) are the same as in the scalar case, except we now have to keep track of the degeneracy in the $a$ index. In principle, we could perform the analysis for any $n$. However, due to the time-consuming nature of such analysis at higher dimensions and higher levels, here we shall only deal with the case of AdS$_4$ (i.e. $d=3$ or $n=1$).

\subsubsection{$d=3$}
At level zero, there are two states (since $2^n=2$).
At level one, since there are three $P_\mu$'s but there is a spinor degeneracy of $2$, there are $2\times 3 =6$  states. Similar counting gives degeneracy at a general level $k$. Up to level four, the eigenvalues of the inner-product matrix are:
\be
\begin{array}{|c | l |c |}
\hline
k& \text{Eigenvalues} & \text{Multiplicity}\\
\hline
  0 & 1 & 2 \\
\hline
  1 & h-1 & 2 \\
\hline
   & h+\half  & 4 \\
\hline
  2 & (h+\half)(h-1) & 6 \\
\hline
   & (h+\half)(h+\frac{3}{2})  & 6 \\
\hline
  3 &  ( h +\half) h(h -1)&2 \\
\hline
   &( h +\half) ( h +\frac{3}{2}) ( h +\frac{5}{2})&8 \\
\hline
   &(h+\half)(h+\frac{3}{2})(h-1)&10 \\
   \hline
  4 & (h+\half)(h+\frac{3}{2}) h(h-1)    &6\\
\hline
   &  ( h+\half) ( h+\frac{3}{2}) ( h+\frac{5}{2}) ( h+\frac{7}{2}) &10\\
\hline
   &  ( h+\half) ( h+\frac{3}{2})  (h+\frac{5}{2})(h-1)  &14\\
\hline
\end{array}
\ee 
So for special values of $h$ where the representation is shortened and finite, the degeneracy is:
\be
\begin{array}{|c | c|}
\hline
-h& \text{Dimension} \\
\hline
  \half & 2+2=2\times(1+1)=4 \\
    \hline
  \frac{3}{2} & 2+2+4+6+2=2\times(1+3+3+1)=16 \\
    \hline
  \frac{5}{2} & 2+2+4+6+6+2+10+6+2=
2\times(1+3+6+6+3+1)
=40\\
    \hline
    \end{array}
\ee These results suggest the pattern that finite representations occur whenever
\be
h=-\half - p
\ee with
 the degeneracy of
 \be
D(p)= 2^2\times\left[1+3+6+10+ \ldots+\half (p+1)(p+2)\right]
=\frac{2}{3}(p+1)(p+2)(p+3) \,.
\ee


\bibliography{QNMbib2}
\end{document}